\documentclass[11pt]{article}
\usepackage{moriond,epsfig}
\epsfclipon

\def\be{\begin{equation}}
\def\ee{\end{equation}}
\def\bea{\begin{eqnarray}}
\def\eea{\end{eqnarray}}

\newcommand{\rsig}{\ensuremath{R_\sigma}}

\newcommand{\sll}{\ensuremath{\sigma_{\ell\ell}}}
\newcommand{\slj}{\ensuremath{\sigma_{\ell j}}}
\begin{document}
\vspace*{4cm}
\title{TOP PHYSICS:  CDF RESULTS}

\author{ KENNETH BLOOM \\
for the CDF Collaboration }

\address{Department of Physics, University of Michigan,\\
Ann Arbor, MI 48109, USA}

\maketitle\abstracts{ The top quark plays an important role in the
grand scheme of particle physics, and is also interesting on its own
merits.  We present recent results from CDF on top-quark physics based
on 100-200 pb$^{-1}$ of $p\bar{p}$ collision data.  We have measured
the $t\bar{t}$ cross section in different decay modes using several
different techniques, and are beginning our studies of top-quark
properties.  New analyses for this conference include a measurement of
$\sigma_{t\bar{t}}$ in the lepton-plus-jets channel using a neural net
to distinguish signal and background events, and measurements of
top-quark branching fractions.}

Evidence for the top quark was first reported ten years ago this
spring~\cite{bib:NYT}, but the properties of top, a fundamental
particle in the standard model, have yet to be fully determined.
There is still much to learn about the processes of production and
decay, and we have not yet measured the quantum numbers of top, which
are exactly predicted by the standard model.  These properties are
important not just for the purpose of characterizing top, but also for
understanding its relationship to other particles in the model.
Predictions for the mass of a standard-model Higgs boson depend very
strongly on the value of the top-quark mass~\cite{bib:gambino}.
Because top is so heavy, it may provide a path to new physics.  The
top quark must be understood in detail so that we can understand
whether it is just part of the standard model, or the first hint of
physics beyond it.

The top quark is an active topic of study in the CDF Collaboration,
which operates a detector that records data from $p\bar{p}$ collisions
at $\sqrt{s} = 1.96$~TeV.  The upgraded detector was commissioned in
2001, and data-taking for physics began in 2002.  Since then, CDF has
recorded about 200~pb$^{-1}$ of analysis-quality data in Run~II, roughly
double the sample of the previous data run.  Here we present
measurements of top-quark production rates and properties in this data
sample.

Most of our studies are done with $t\bar{t}$ pairs, which are produced
through the strong interaction.  We typically assume that $B(t\to Wb)
= 100\%$.  Every $t\bar{t}$ event thus has two $W$'s, so we classify
the final states by the decays of the $W$'s.  ``Dilepton'' events have
both $W$'s decaying to $\ell\nu$; this mode has good signal to noise
($S/N \sim$~1.5-3.5), but a low detectable event rate (4-6
events/100~pb$^{-1}$).  ``Lepton plus jets'' events have only one $W$
decaying to $\ell\nu$; the rate is higher than for dileptons (25-45
events/100~pb$^{-1}$) but this channel has worse $S/N$
($\sim$~0.3-3.0).  Regardless of how the $W$ decays, $t\bar{t}$ events
have two $b$ quarks; these can be identified by displaced vertices
(due to the long $b$-hadron lifetime and large mass) or soft leptons
(due to the large $b$-hadron semileptonic branching fraction) embedded
in the jets produced by quark hadronization.

\section{Cross-Section Measurements}
Before we can study the properties of the top quark, we have to
isolate a sample of candidate top-quark events, and build tools to
understand the rates of background events in the sample and the
efficiency for top-quark selection.  Our benchmark for these tools is
a measurement of the $t\bar{t}$ production cross section.  This
production rate is predicted by QCD; a measurement that agrees with
theory gives us confidence that we have some control over our tools,
and that we in fact have top quarks to study.

The cross-setion measurement for dileptons is straightforward because
the events are so clean and the backgrounds are relatively small.  CDF
has two complementary dilepton cross-section measurements.  One
explicitly identifies the two leptons (electrons or muons), while the
other identifies only one lepton and searches for an additional
isolated high-$p_T$ track.  The former analysis has lower background,
but the latter has greater acceptance and greater sensitivity to
$\tau$ decays of the $W$.  The lepton-lepton analysis observes 13
events (1 $ee$, 3 $\mu\mu$, 9 $e\mu$) on $2.4 \pm 0.7$ background
events, while the lepton-track analysis observes 19 events (11
$e$-track, 8 $\mu$-track) on $7.1 \pm 1.2$ background -- significant
signals in both cases.

\begin{figure}
\psfig{figure=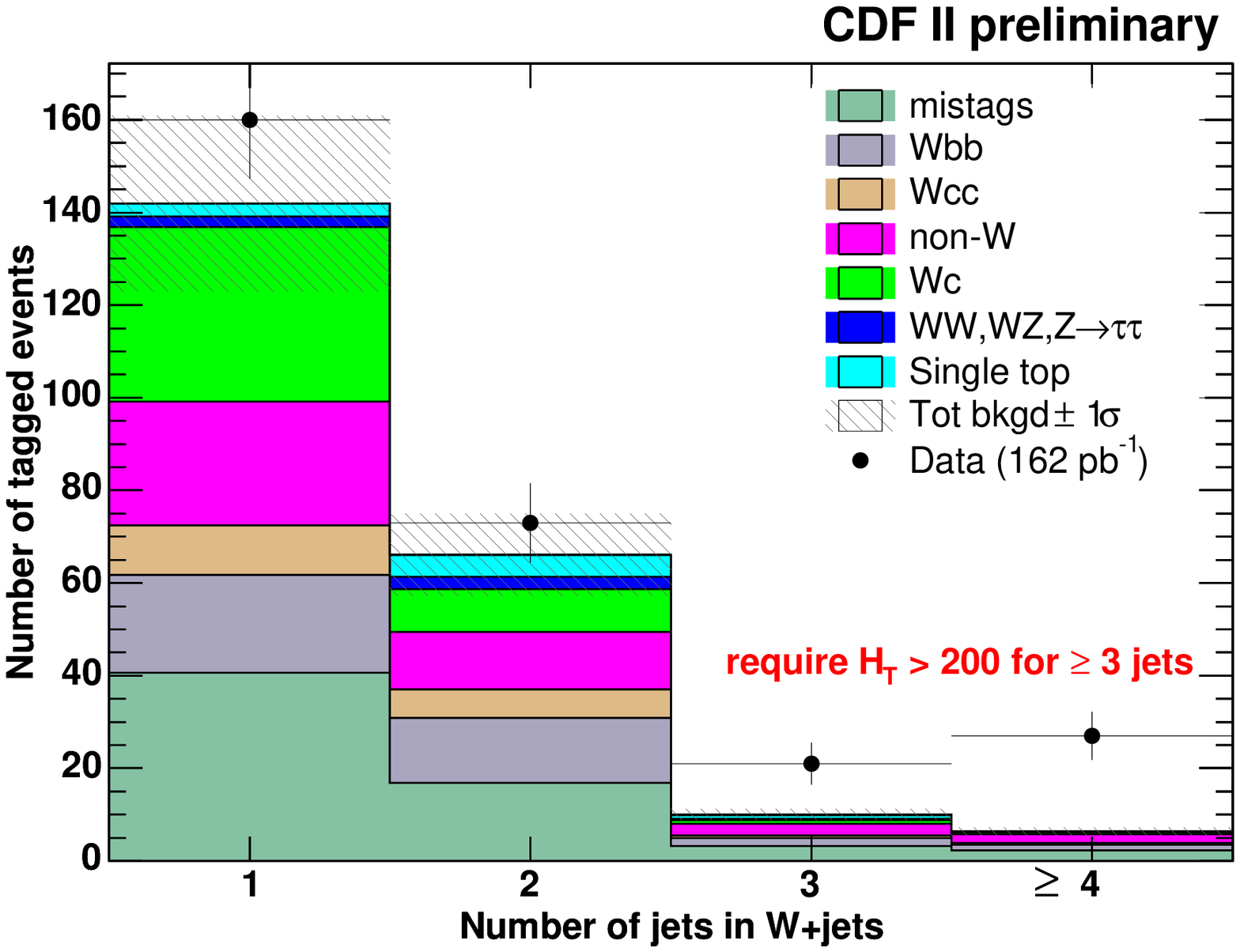,height=2.55in}
\psfig{figure=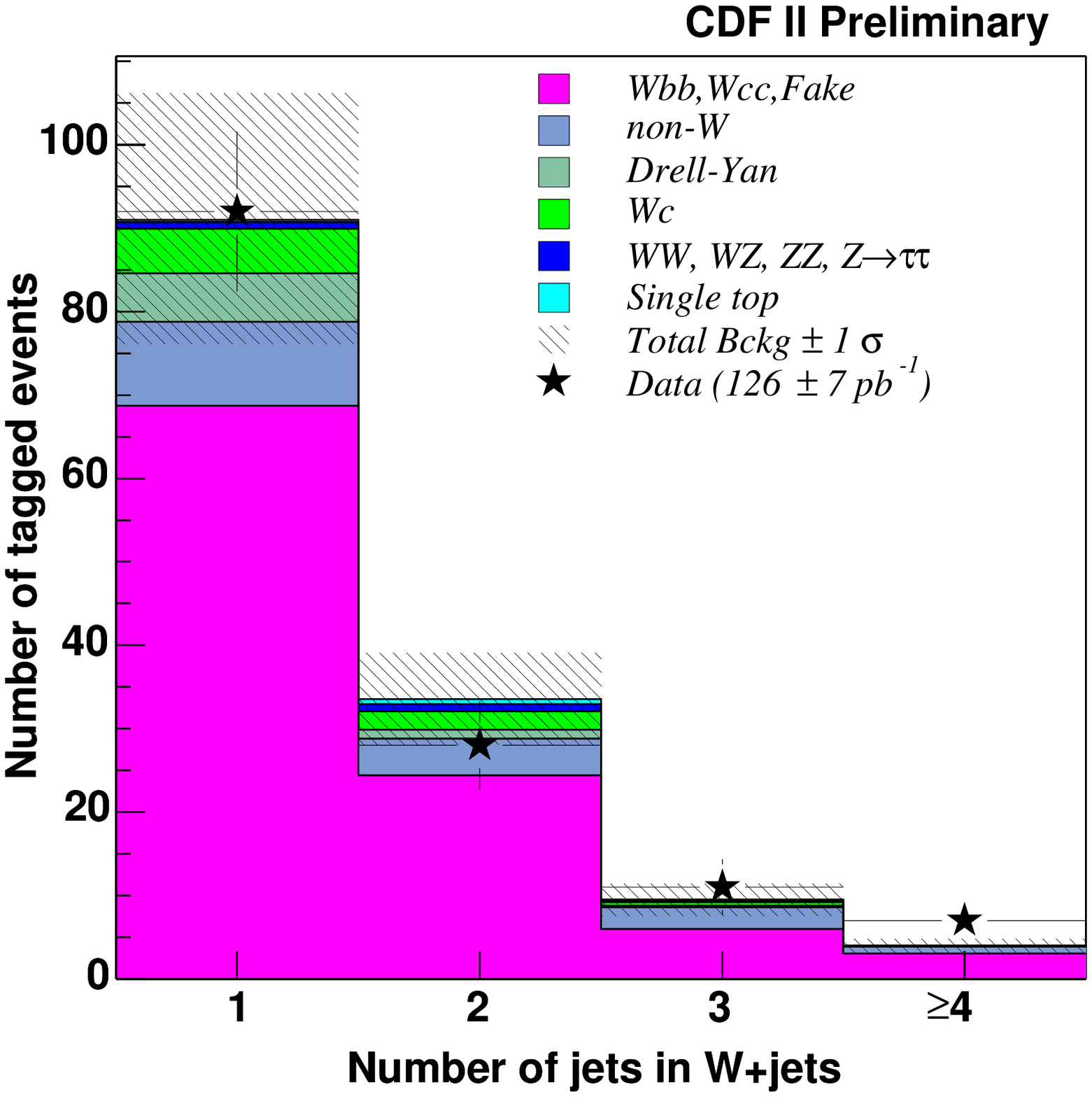,height=2.55in}
\caption{$N_{jet}$ distributions in lepton-plus-jets samples which
contain vertex tags (left) and lepton tags (right), along with
estimates of contributions of standard model non-$t\bar{t}$ processes.
\label{fig:njet}}
\end{figure}

Cross-section measurements in the lepton-plus-jets mode fall into two
categories.  The first includes counting experiments, in which we
predict from first principles the rate of standard-model
non-$t\bar{t}$ backgrounds in the sample; any excess over that rate is
considered the $t\bar{t}$ signal.  Such counting experiments are only
done in samples with at least one tagged $b$ jet, where the
non-$t\bar{t}$ backgrounds are relatively small.  The background
estimates are data-driven as much as possible; estimates for rates of
QCD-jet and fake-tag backgrounds are entirely derived from data, while
backgrounds from $W$ plus heavy-flavor processes ($Wb\bar{b}$,
$Wc\bar{c}$, $Wc$) and lower-rate processes (dibosons, single top) get
assistance from Monte-Carlo information.  After checking that these
methods work in lepton-plus-jets samples with fewer than three jets,
where we expect little $t\bar{t}$, we apply them to the
lepton-plus-jets samples with at least three jets, where most of the
$t\bar{t}$ events are expected.  The distribution of the number of
jets in $b$-tagged events is shown in Figure~\ref{fig:njet}.  We use
two different $b$-tag techniques.  In events where there is at least
one jet with a displaced-vertex tag, we observe 57 events with a
background prediction of $23.4 \pm 3.0$.  In events where there is at
least one jet with a lepton tag, we observe 18 events with a
background prediction of $12.9 \pm 2.6$.

\begin{figure}
\psfig{figure=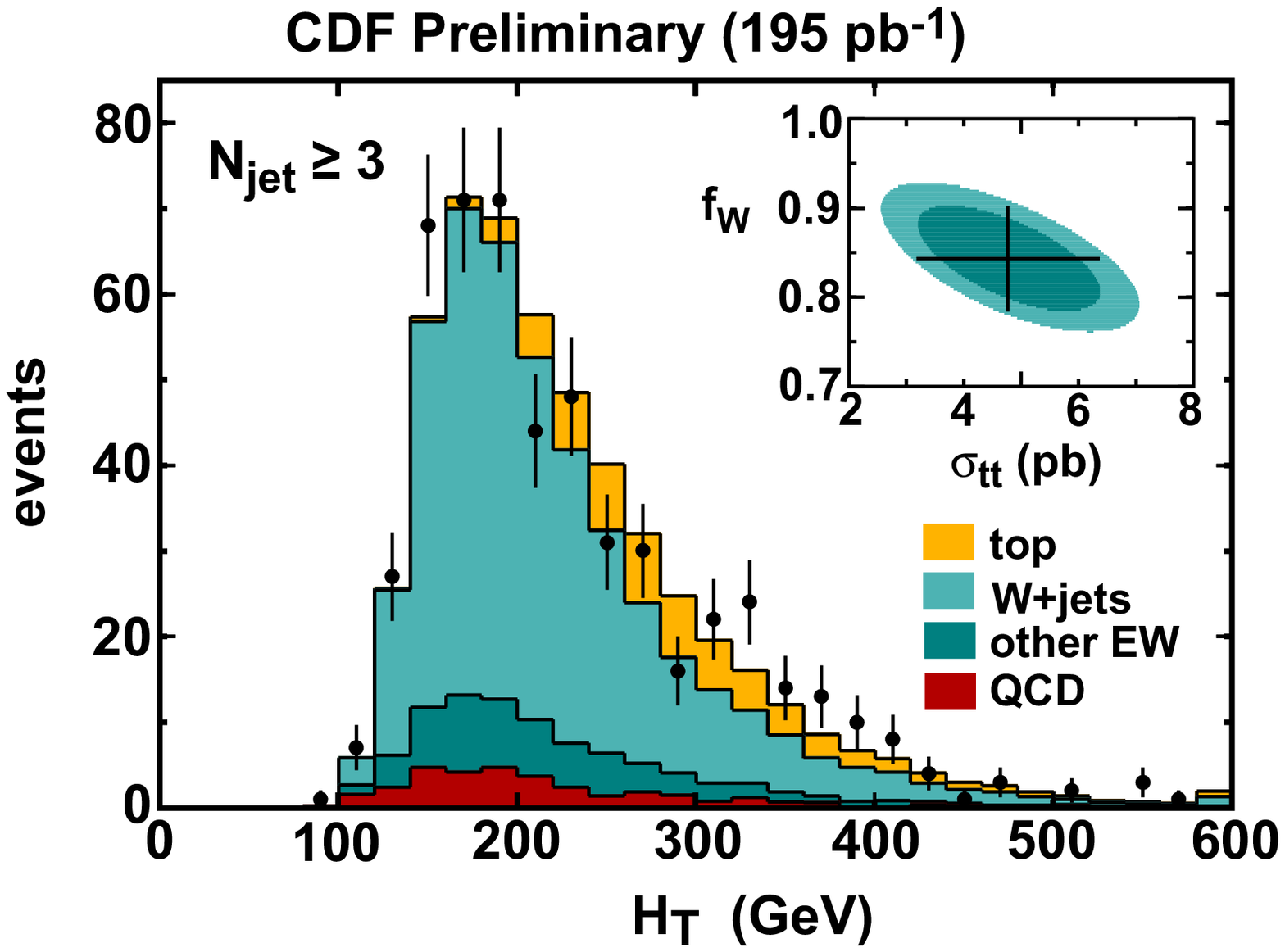,height=2.6in}
\psfig{figure=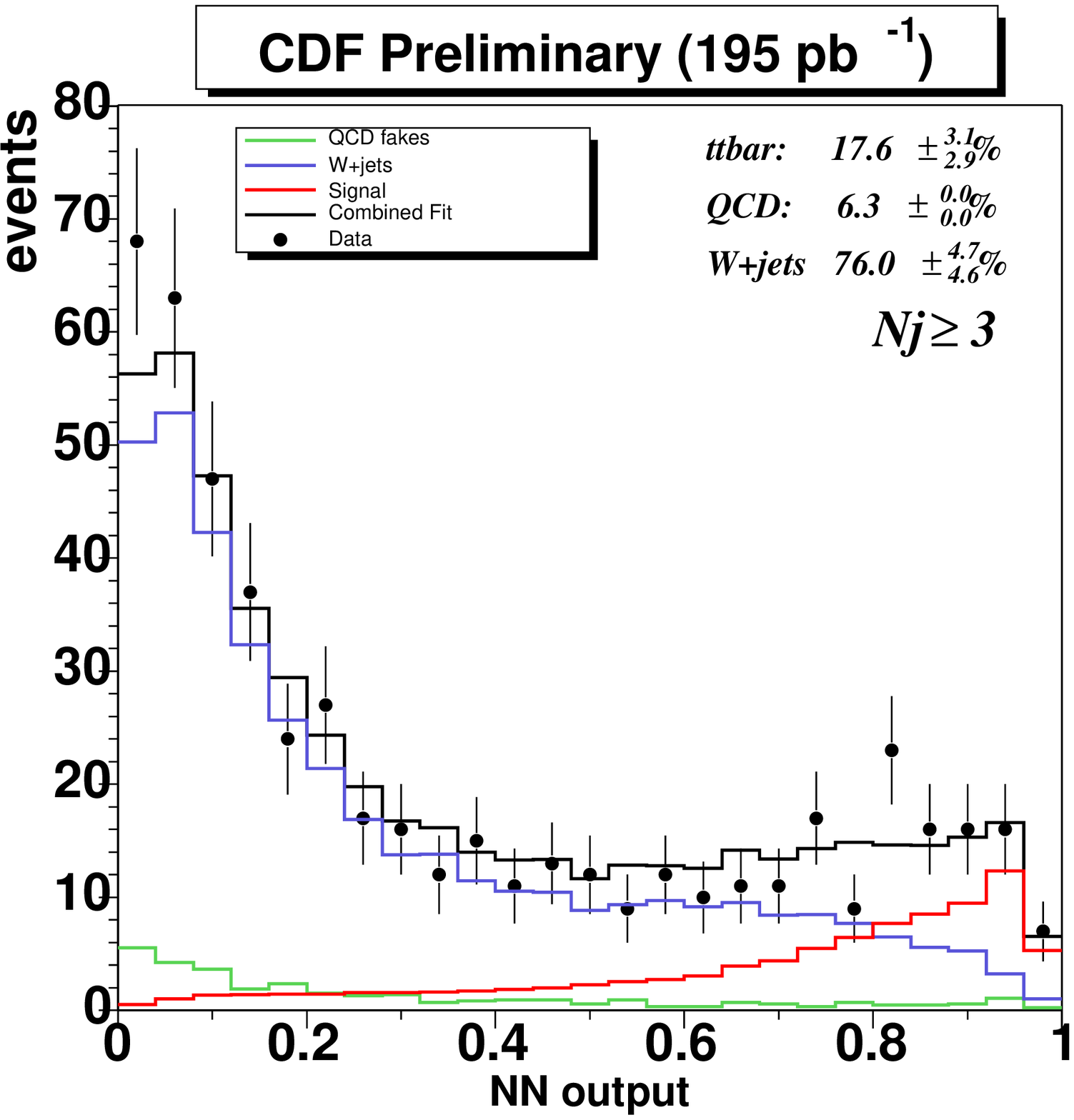,height=2.6in}
\caption{Distributions of $H_T$ (left) and the neural-network output
(right) in the lepton-plus-jets sample (without requiring $b$ tags),
with models of the physics processes that contribute to this sample.
\label{fig:htnn}}
\end{figure}

Cross-section measurements that make use of kinematic information in
candidate events are in the second category.  Rather than
dead-reckoning the background rate, we use characteristics of the
signal sample itself to help estimate the rate.  We build models of
kinematic variables for signal and background events, and then fit the
distributions observed in the data to a sum of these models, letting
the normalizations float.  The estimated fraction of 
$t\bar{t}$ events can be converted into a cross section.  Such
analyses can be done with lepton-plus-jets samples where $b$ tags are
not required, as we use the kinematic information to help separate
signal from background instead of the tags.  This exploits the large
event sample, but now the backgrounds to $t\bar{t}$ are larger.  With
a fit to the distribution of the total transverse energy of the event
($H_T$, see Figure~\ref{fig:htnn}), we find that this sample has a
($13 \pm 4$)\% $t\bar{t}$ content.  The signal and background $H_T$
shapes are modeled by Monte Carlo simulation, and systematic effects
such as our understanding of the jet-energy scale and inputs to the
leading-order matrix-element Monte Carlos (such as the $Q^2$ scale)
limit our precision.  A complementary approach is to perform this
measurement in the sample of lepton-plus-jets events that include
displaced-vertex-tagged jets; the energy of the leading jet is used as
the kinematic variable.  Now the signal-to-noise is improved, and we
can use a sample of real data to model the background -- the
lepton-plus-jets events without $b$ tags, which should be depleted of
$t\bar{t}$ events -- and thus largely eliminate the systematic
uncertainty due to modeling.  The $t\bar{t}$ content of this sample is
($67^{+13}_{-16}$)\%, where the uncertainty is now dominated by
statistics.

While the kinematics of $t\bar{t}$ and the dominant $W$ plus jets
background process do differ, they only differ modestly, so there is
very little separation in any given variable.  However, these modest
differences occur in many different variables.  Thus, we choose
several variables (jet energies, invariant masses, event shapes) that
are largely uncorrelated, or have different correlations in signal and
background processes, and then develop a neural network from these
variables to use the information optimally.  The distribution of
neural-net outputs for the lepton-plus-jets sample (the same sample
used in the $H_T$ analysis) is also shown in Figure~\ref{fig:htnn}.
Now there is much better separation between the $t\bar{t}$ signal and
the backgrounds, which leads to improved statistical and systematic
uncertainties.  The $t\bar{t}$ fraction in this sample is ($18 \pm
3$)\%, consistent with, and more accurate than, the value measured in
the $H_T$ analysis.  We believe that such multivariate techniques are
very promising for isolating $t\bar{t}$ and other hard-to-extract
signals.

\begin{figure}[ht]
\centering
\psfig{figure=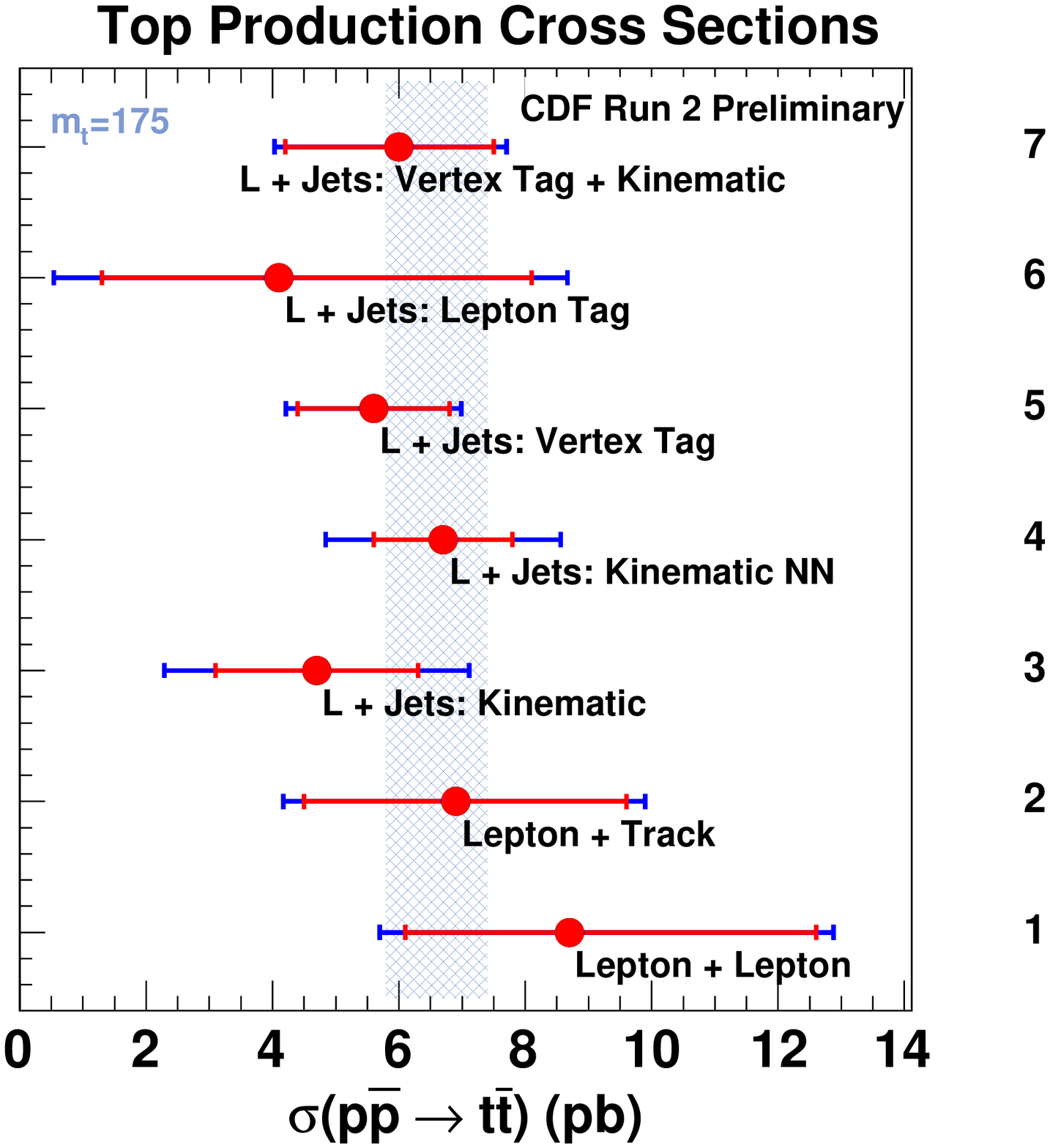,height=4.9in}
\caption{Summary of $t\bar{t}$ cross-section measurements from CDF;
the shaded band is the theory prediction.
\label{fig:xsec}}
\end{figure}

A summary of the current set of $t\bar{t}$ cross-section measurements
from CDF is given in Table~\ref{tab:xsec} and Figure~\ref{fig:xsec},
along with the current prediction from theory~\cite{bib:xsectheory}
with $m_t = 175$~GeV.  One should not perform a na\"{\i}ve average of the
values as there are correlations, and one should remember that the
theory prediction has an additional $\sim$1~pb uncertainty due to the
uncertainty in $m_t$, but we do see general consistency among the
measurements, and agreement with the theory prediction.  This gives us
confidence that we are in fact observing $t\bar{t}$ production.

\begin{table}[t]
\caption{Summary of $t\bar{t}$ cross-section measurements from CDF.
The first uncertainty is statistical, the second is systematic.
\label{tab:xsec}}
\vspace{0.4cm}
\begin{center}
\begin{tabular}{|c|c|c|}\hline
Analysis & $\sigma_{t\bar{t}}$~(pb) & $\int{\cal L} dt$~(pb) \\\hline
Dilepton -- lepton-lepton & $8.7^{+3.9}_{-2.6} \pm 1.5$ & 193 \\
Dilepton -- lepton-track & $6.9^{+2.7}_{-2.4} \pm 1.3$ & 200 \\
Lepton + jets -- kinematic ($H_T$) & $4.7 \pm 1.6 \pm 1.8$ & 195 \\
Lepton + jets -- kinematic (NN) & $6.7 \pm 1.1 \pm 1.5$ & 195 \\
Lepton + jets -- vertex tag & $5.6 \pm 1.2 \pm 0.7$ & 162 \\
Lepton + jets -- lepton tag & $4.1^{+4.0}_{-2.8} \pm 2.2$ & 126 \\
Lepton + jets -- vertex tag + kinematic & $6.0^{+1.5}_{-1.8} \pm 0.8$ & 162 \\
\hline
Theory ($m_t = 175$~GeV)~\cite{bib:xsectheory} & $6.7^{+0.6}_{-0.9}$ & \\\hline
\end{tabular}
\end{center}
\end{table}

In addition, CDF is searching for the production of single top quarks
through the electroweak interaction; the cross section is expected to
be about 3~pb.  Eventually this will provide a direct measurement of
$|V_{tb}|^2$.  We currently set a 95\% CL upper limit of 13.7~pb on
the total single-top production rate. 

\section{Branching fractions}
With the top-quark sample identified, we can begin to measure the
properties of top quarks.  For instance, we can use the cross-section
measurements to study the branching fractions of top.  The standard
model predicts that $B(t\to Wb)$ is virtually 100\%; in particular, we
expect that there is always a $W$ in the decay.  The $t\bar{t}$
cross-section measurements in fact assume that there are always two
$W$'s in the final state; the event rates are corrected for the $W$
branching fractions to obtain the total $t\bar{t}$ cross section.

We can test this assumption by examining the ratio of measured cross
sections for different $t\bar{t}$ final states, $\rsig = \sll/\slj$.
$\rsig$ should be consistent with unity if in fact the dilepton and
lepton-plus-jets analyses are both examining standard-model $t\bar{t}$
production.  Measuring ratios is appealing to experimenters; $\rsig$
will have smaller systematic uncertainties than the individual
cross-section measurements as some common factors will cancel.  And of
course, $\rsig$ is independent of any theory prediction for
$\sigma_{t\bar{t}}$, so we can look for new physics by looking for a
deviation in the ratio, rather than by comparing a measured cross
section to an uncertain theory prediction.  As an example, $\rsig$ is
sensitive to decays such as $t\to H^+ b$.  If this process occurs,
then the mix of dilepton and lepton-plus-jets events would be
different from what is expected in the standard model, and that mix is
sensitive to $\tan\beta$, which controls whether the $H^+$ is more
likely to decay to hadrons or to leptons.

\begin{figure}
\hspace{0.5in}
\psfig{figure=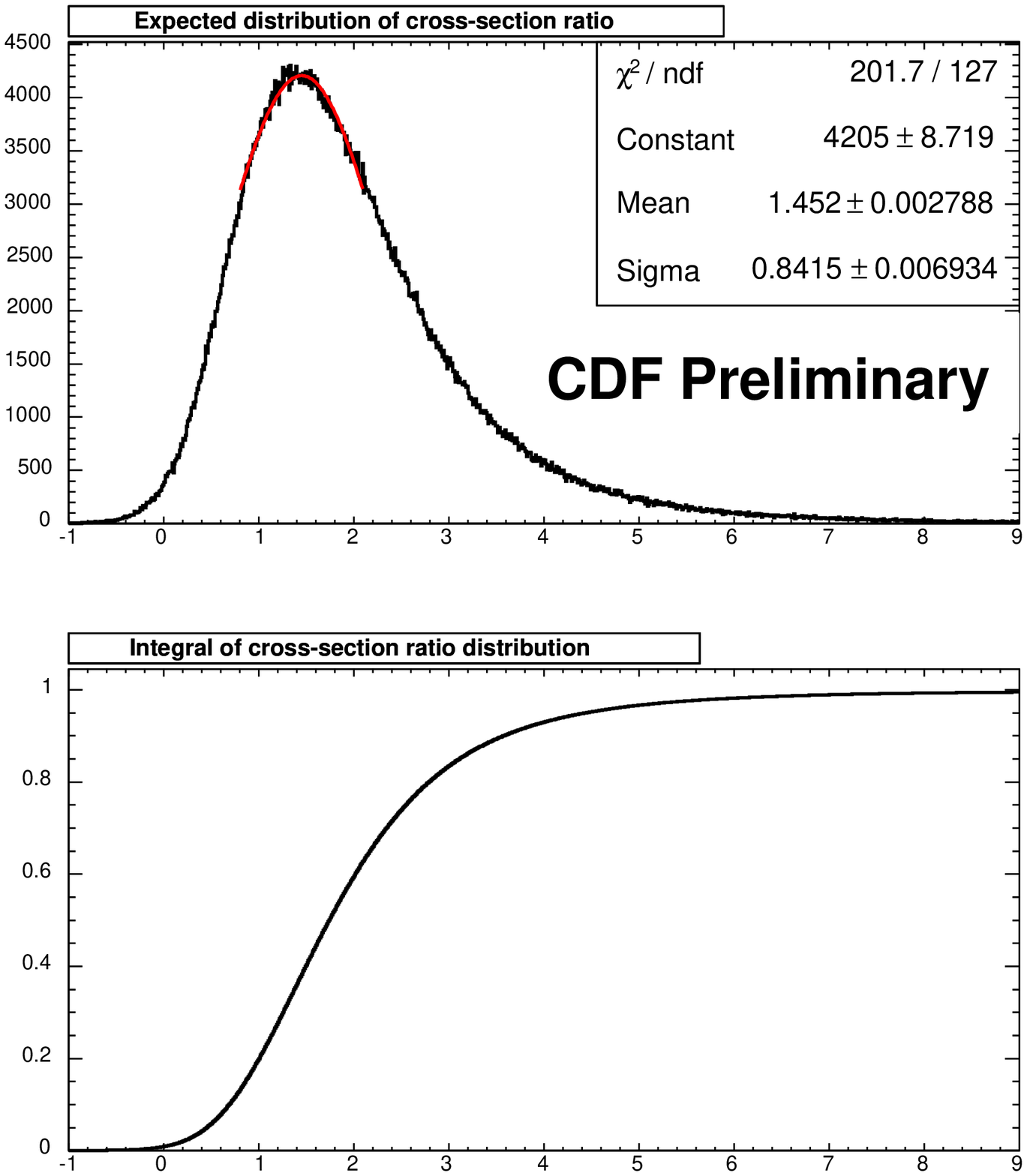,height=3.0in}
\caption{Probability distribution for $\rsig = \sll/slj$.
\label{fig:ratio}}
\end{figure}

We estimate $\rsig$ creating a probability distribution based on the
observed event rates; this is shown in Figure~\ref{fig:ratio}.  We
find $\rsig = 1.45^{+0.83}_{-0.55}$, and we can limit $0.46 < \rsig <
4.45$ at 95\% CL.  Since this is consistent with standard-model
expectations, we can set limits on non-standard decays of top.  These
limits are by their nature model dependent, as we need to understand
the efficiency to detect the non-standard decay.  We find that the
branching fraction to an all-hadronic $t\to Xb$ decay is less than
0.46 at 95\% CL under the assumption that we detect standard and
non-standard all-hadronic decays with the same efficiency.

We can also examine the other side of the branching-ratio coin -- we
typically assume that there is always a $b$ in each top decay, and in
fact all of the cross-section measurements that use $b$ tagging depend
on that assumption.  We can test that hypothesis by examining the
$b$-tag rates in identified $t\bar{t}$ events.  These rates depend
both on $b = B(t\to Wb)/B(t\to Wq)$ and on the single-$b$ tagging
efficiency $\epsilon$.  The rates of observing two, one or zero tags
in $t\bar{t}$ events depend on the product $b\epsilon$, as the $b$
quark must be produced and subsequently tagged:
\begin{eqnarray*}
N_2 \propto (b\epsilon)^2 & N_1 \propto 2b\epsilon(1 - b\epsilon) &
N_0 \propto (1 - b\epsilon)^2 \Rightarrow
\end{eqnarray*}
\begin{eqnarray*}
b\epsilon = \frac{2}{N_1/N_2 + 2} = \frac{1}{2 N_0/N_1 + 1}.
\end{eqnarray*}
$b\epsilon$ is thus determined by the ratios of tag rates.  Since we
measure the product $b\epsilon$, we can either take the value of
$\epsilon$ measured in calibration samples and extract $b$, or just
assume the standard-model value $b=1$ and do an {\it in situ} cross
check of $\epsilon$.

\begin{figure}
\hspace{0.5in}
\psfig{figure=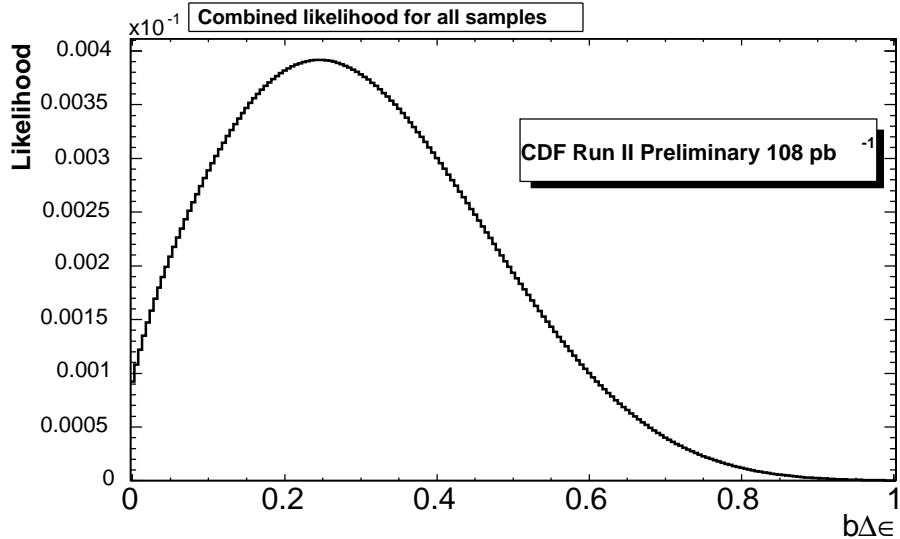,height=3.0in}
\caption{Likelihood distribution for $b\epsilon$ in vertex-tagged
lepton-plus-jets sample.
\label{fig:twb}}
\end{figure}

The above treatment is simplistic; we need to account for our limited
acceptance for $b$ jets, for tagged jets that come from non-$t\bar{t}$
events, and for tagged jets that come from non-$b$ quarks in
$t\bar{t}$ events.  We create a likelihood function that depends on
the observed numbers of tags in the vertex-tagged lepton-plus-jets
sample, and find the most likely value of $b\epsilon$.  The resulting
likelihood is shown in Figure~\ref{fig:twb}.  We measure the most
likely value as $b\epsilon = 0.25^{+0.22}_{-0.18}$.  The single-$b$
tagging efficiency is $0.45 \pm 0.05$, and thus $b =
0.54^{+0.49}_{-0.39}$ -- smaller than expected, but consistent with
$b=1$.  If we assume $b=1$, we find an estimate of $\epsilon$ that is
consistent with our measurements in calibration samples.  In addition,
we can set a lower limit of $b > 0.12$ at 95\% CL.  If this analysis
remains exactly the same as we record more data -- no change in
background levels or efficiencies -- we expect to set a lower limit of
about 0.70 with 500~pb$^{-1}$ of integrated luminosity.  Of course, we
do hope to reduce our backgrounds and increase our efficiencies as we
accumulate more data.

\section{Mass}
As stated above, measurements of the top-quark mass are of great interest
right now, as predictions for the mass of the Higgs are so sensitive to it.
As of the date of this conference, CDF has performed two preliminary
measurements of the top mass with Run~II data.

One makes use of the dilepton sample, which has very little background
but low event rates.  Because of the two neutrinos in the final state,
estimating the top mass is an underconstrained problem, and some more
information must be added.  In addition, we do not know which jets to
match up with which leptons in the final state, giving an additional
combinatorics problem.  The current CDF analysis scans over possible
directions of the neutrinos, and uses the predicted $p_T$ of the
$t\bar{t}$ system (taken from theory) as an event weight.  Multiple
solutions for the mass are possible; we choose the one with the
greatest kinematic consistency with $t\bar{t}$.  Using six events in
126 pb$^{-1}$ of data, we measure $m_t = 175 \pm 17_{\rm stat} \pm
8_{\rm sys}$~GeV, where the systematic uncertainty is dominated by our
limited knowledge of the jet-energy scale.

Another mass measurement is made with the lepton-plus-jets sample.  We
choose the events with four jets, at least one of which is
vertex-tagged.  The event rates are larger here than in the dilepton
mode, and while the backgrounds are also larger, they are tolerable.
Now that there is only one neutrino in the final state, the problem is
overconstrained.  In each event, we choose the combination of jets
that gives the best consistency with the $t\bar{t}$ hypothesis in a
kinematic fit.  With 22 events in a sample of 108~pb$^{-1}$, we
measure $m_t = 178^{+13}_{-9 {\rm stat}} \pm 7_{\rm sys}$~GeV.  Again,
the systematic uncertainty is dominated by the jet-energy scale.
Improved measurements of the top-quark mass are expected shortly.

\section{Outlook}
At this stage of Run~II at the Tevatron, there is little doubt of the
top-quark signal at CDF.  We have observed $t\bar{t}$ production in
multiple channels (dilepton and lepton plus jets) and with multiple
techniques (different $b$ taggers, counting experiments, and fits to
kinematic distributions, including a new neural-net-based analysis).
All of the $t\bar{t}$ cross-section measurements are consistent with
the theory prediction, and with each other.

Meanwhile, studies of top-quark properties are beginning.  At this
conference, we presented the first Run~II measurements of top-quark
branching fractions.  We are hard at work on the mass measurements:
adding in more data, improving our techniques and reducing systematics.
We hope to have new results imminently.

This is just the beginning of a broad program to characterize the top
quark.  At CDF, we are preparing for a variety of top-quark studies --
measurements of the $W$ helicity, searches for decays to charged Higgs
and for resonant $t\bar{t}$ production, and studies of angular
correlations in top decay.  More data and more news about top and its
impact on standard and new physics are expected soon.

\section*{Acknowledgements}
I thank my CDF colleagues for all of their efforts to produce these
results, and the conference organizers for putting together a very
pleasant week of physics and other activities.

\end{document}